# A Transformer-Based Approach for DDoS Attack Detection in IoT Networks

Sandipan Dey[1], Payal Santosh Kate[2], Vatsala Upadhyay[3], Dr. Abhishek Vaish[4]

1,2,3,4: Department of Information Technology, Indian Institute of Information Technology Allahabad, Prayagraj-211015, Uttar Pradesh, India.

Corresponding author: Dr. Abhishek Vaish, abhishek@iiita.ac.in

**Abstract:** DDoS attacks have become a major threat to the security of IoT devices and can cause severe damage to the network infrastructure. IoT devices suffer from the inherent problem of resource constraints and are therefore susceptible to such resource-exhausting attacks. Traditional methods for detecting DDoS attacks are not efficient enough to cope with the dynamic nature of IoT networks, as well as the scalability of attacks, diversity of protocols, high volume of traffic, and variability in device behavior, and variability of protocols like MQTT, CoAP, making it hard to implement security across all the protocols. In this paper, we propose a novel approach, i.e., the use of Transformer models, which have shown remarkable performance in natural language processing tasks, for detecting DDoS attacks on IoT devices. The proposed model extracts features from network traffic data and processes them using a self-attention mechanism. Experiments conducted on a real-world dataset demonstrate that the proposed approach outperforms traditional machine learning techniques, which can be validated by comparing both approaches' accuracy, precision, recall, and F1-score. The results of this study show that Transformer models can be an effective solution for detecting DDoS attacks on IoT devices and have the potential to be deployed in real-world IoT environments.

## 1 Introduction

The rapid growth of home automation systems has provided numerous entry points for hackers to exploit. Internet-connected devices such as smart refrigerators, smart watches, smart fire alarms, smart door locks, medical sensors, fitness trackers, wireless lamps, surveillance equipment, and smart thermostats with attached sensors offer hackers significant opportunities to access these systems. While IoT has made daily tasks easier to manage, it is needed to ensure that hackers cannot take advantage of these vulnerabilities[1]. As the amount of data generated by these devices increases with the lightning growth of the 5G network, it is essential to ensure that the data is secured to prevent it from being stolen for financial gain or potentially risking people's lives[2]. In 2018, around 50% of the most significant global exploits targeted IoT devices, with IP cameras accounting for the majority of these attacks[3]. Even defense mechanisms like cameras are susceptible to such attacks. With 2.98 billion cyber-attacks registered in the first half of 2019, the number of cyber-attacks on IoT devices surged more than three times compared to the second half of 2018[4]. In 2020, there were about 30 billion IoT devices online; by 2025, that number is expected to rise to 75 billion[5]. Evidence of this need can be seen in the fact that investments in IoT security climbed by more than 25% to reach 2.5 billion US dollars in 2020. This shows the need to outfit IoT devices with adequate security[6] appropriately. One such vulnerability is Distributed Denial of Services (DDoS), which employs compromised IoT devices to attack their targets by infecting them with a Botnet[53],[54]. Flooding threats known as distributed denial of service (DDoS) prevent genuine users from using the service they have paid for. According to current market data, it is one of the most common cybersecurity threats. Therefore, it is crucial to recognize and comprehend the various DDoS threats[7], [54]. There are primarily two types of DDoS attacks: reflection-based and exploitation-based. Each DDoS attack fits into one of these two categories. Reflection-based DDoS attacks are ones in which the hacker conceals his identity by utilizing tools and components from a third party[8]. Data packets, including the source and IP address of the victim or target, are sent to a reflector server to start the attack. These attacks are carried out using the User Datagram Protocol (UDP), Transport Control Protocol (TCP), or both simultaneously through the Application layer protocol. NTP, TFTP, MSSQL, and SSDP are examples of TCP-based attacks.DNS, LDAP, NETBIOS, and SNMP are examples of TCP/UDP-based combination attacks[9]. Reflection-based and Explosion-based attacks require third-party software and components to maintain anonymity when the attack is launched. It has both TCP- and UDP-based attacks. SYN Flood is a component of TCP-based attacks, while UDP-based attacks include UDP-Flood and UDP-lag. The hacker starts the UDP threat by delivering a massive number of UDP packets at a very fast rate on the target machine's random ports, which causes the bandwidth to run out and ultimately causes the system to crash. UDP lag slows down users by interfering with client-server connections, which are handled by hardware switches or software that uses up network capacity. The TCP three-way handshake, which includes sending SYN packets continually until the machine collapses, is exploited in SYN attacks[10].

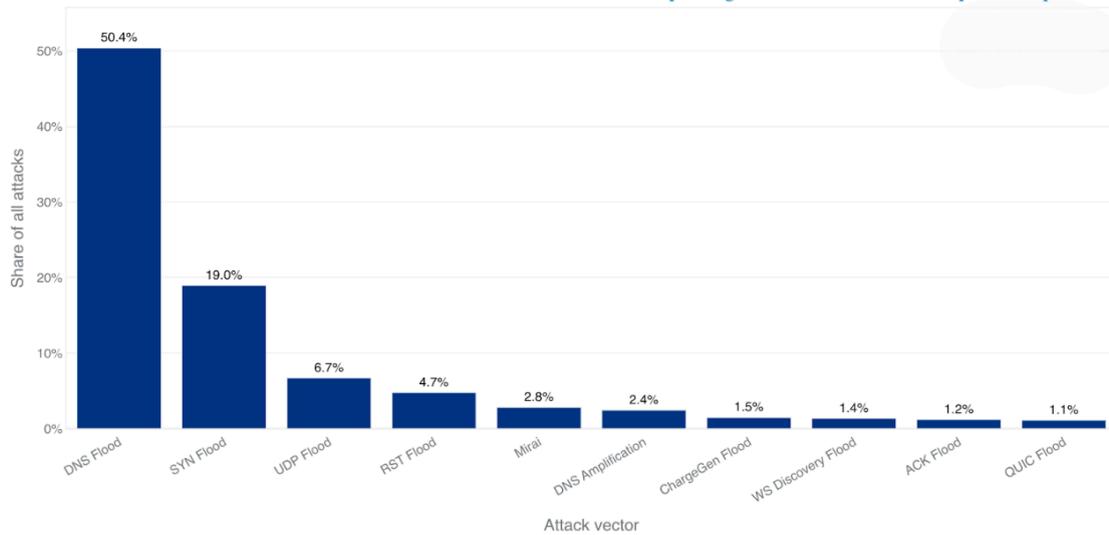

Fig. 1: Top attack vectors in Quarter 4 2023

Figure 1, which was derived from a recent post by Cloudflare about 2023, illustrates that DNS Flood attacks are the most prevalent, making up over half of the recorded network-layer DDoS attacks, followed by SYN Flood attacks, which constitute nearly a fifth of the attacks. According to research by Krebson security [11], each DDoS attack may have cost the owners of the affected devices $323,973.75 in addition to additional costs for the excessive use of power and bandwidth. The severity of the situation makes it one of the main issues with internet security, and several statistical detection techniques, such as wavelet-based, port entropy-based, destination entropy-based, etc., can be discovered in the various literature. However, given the constantly evolving nature of the internet, these techniques are often very time-consuming. As a result, several researchers have explored using machine learning and artificial intelligence techniques to identify DDoS attacks as a solution to these problems[46], [51]. Finding the most recent and accurate DDoS data set that correctly reflects IoT-specific behavior is the most challenging issue in developing a machine learning model to identify DDoS attacks [11], [47]. UNSW-NB15 and BoT-IoT are the most well-known and recent datasets in the area that encompass IoT devices using both simulated and actual data [12]. For tasks involving network security, such as DDoS detection, deep learning models are widely used [13], [49], [52]. Given its ability to spot long-term connections in sequential data, the Transformer architecture, one of these models, has garnered a lot of attention [14]. Moreover, transformer-based methodologies use parallel processing, enabling them to capture all the long-range dependencies within less time than traditional ML models [15]. Also, it overcomes the scalability issue; the traditional methods provide good results for small-scale IoT networks, but when the network is magnified, the accuracy drops significantly [16]. DDoS attacks usually involve a significant amount of dynamically changing network traffic, making it a strong option for detecting them [17], [48], [50]. [50] addresses the critical issue of communication security within the Internet of Things (IoT) ecosystem, focusing on preventing Distributed Denial of Service (DDoS) attacks. The authors analyze the security challenges inherent in IoT networks and propose a lightweight defensive algorithm designed to mitigate the risks of DDoS attacks. The paper emphasizes the need for effective security measures due to the widespread deployment of IoT devices, which are often susceptible to various cyber threats. [46] presents a novel, time-efficient approach for detecting Distributed Denial of Service (DDoS) attacks in Internet of Things (IoT) networks by leveraging Software-Defined Networking (SDN). The authors focus on addressing the challenges of real-time DDoS detection in IoT environments, where traditional detection mechanisms may be ineffective due to the resource constraints and dynamic nature of IoT devices. The paper proposes a hybrid detection model that integrates signature-based and anomaly-based detection methods to enhance the accuracy and efficiency of DDoS detection. [18], [19], [20] These papers justify through their results the benefits of using the Transformer model in IoT networks. Looking into the several benefits of the transformer model for detecting DDoS attacks, we aim to use a DDoS detection-based model. Utilizing an existing learned language model, our approach changes it to match the network traffic data of IoT devices. We compare our approach to other cutting-edge DDoS detection methods using a real-world data set. The results of the studies indicate the efficacy of our strategy, giving us more accuracy and fewer false positives when compared to other methods. The computational analysis supplements the result and highlights the practicality of the proposed system. This study helps to develop reliable and efficient DDoS detection systems for IoT devices, which would increase the security of IoT networks. The following is the list of our prominent contributions to this paper:

• Comprehensive Dataset Fusion: The study combines UNSW-NB15 and BoT-IoT datasets, providing a diverse and comprehensive dataset for training and evaluation.
• Advanced Transformer-Based Model: Utilizes a state-of-the-art Transformer model for DDoS attack detection, demonstrating significant improvements over traditional methods.
• Efficacy of Transformer model: Comparison of various Transformer models for accuracy, F1 score, type of classification head, and encoding technique.
• Computational Analysis: Comparison based on LSTM and Transformer model computational parameters.

## 2 Research Gaps

The trend of the contribution of the past research has already been discussed. In this section, the discussions are based on the shortcomings in past work from the perspective of the use of diverse and versatile datasets for the detection of DDoS in IoT environments; the basic assumption is that the DDoS solution should be scalable and can handle the sophisticated attacks and complex DDoS attack and therefore a careful review of the existing work in terms of the dataset, no of features and percentage of training and its performance has been evaluated, that has culminated in the motivation of the current research study. As we have used a combination of UNSW-NB15 and BoT-IoT datasets, the discussions below include past work on these datasets. Koroniotis et al. [32] created the BoT-IoT dataset and tested it using SVM, RNN, and LSTM. It has achieved the best accuracy of 99.9% for SVM, but has considered only a single dataset. As the limitation in the traditional approaches is related to the detection of attacks in the diversity of protocols, dynamic devices' behavior, etc., the new techniques like the Transformer model would give an optimized solution only if the dataset used is versatile with diverse characteristics. Therefore, considering a single dataset is considered a research gap from the optimized detection perspective. Popoola et al. [33] used SMOTE along with DRNN on the BoT-IoT dataset to reduce the imbalance in the dataset, but considered only 5% of the dataset. Similarly, Churcher et al. [34] have implemented various ML methods like KNN, SVM, DT, NB, random forest, ANN, and LR on the BoT-IoT dataset and got an accuracy of 99% for KNN, considering only 2% of the dataset. It is important to observe that such a limitation has significant concerns about its relevance in real-time detection; the accuracy might be higher; however, due to limitations of the dataset, the features of the attacks are not sufficient as a scalable solution, and therefore, are considered as one of the drawbacks in existing research work. Shafiq et al. [35] have developed a new algorithm CorrAUC for feature selection using the area under the curve for a particular ML model; even though the accuracy yielded is 99.99%, the author has used only the best 5 features train the model, using fewer features with high accuracy indicated less scalable approach, especially with new and sophisticated types of attacks, also it indicates the presence of less sub-class of attack vector which suffers with the drawback of low detection rate with sophisticated attacks, the similar limitation is observed in Khraisat et al. [36] using a hybrid IDS combining C5 and SVM classifier and selected only 13 features of the BoT-IoT dataset. Ibitoye et al. [37] used FNN and SNN for DDoS detection, which was trained on 20% of the BoT-IoT dataset. Guizani et al. [38] have incorporated RNN-LSTM on 10% of the UNSWNB15 dataset. Dai et al. [39] used a combination of CNN, Bi-LSTM, and attention mechanism on NSL-KDD, UNSWNB15, and CIC-DDoS2019 datasets, but the authors did not combine these datasets, leading to more prediction time. Similarly, Alkadi et al. [40] have implemented LSTM methodology on the BoT-IoT and UNSW-NB15 datasets separately; also, the computational cost to implement the model on such a resource IoT network was not discussed.

To the best of our knowledge, none of the studies have worked on finding the effectiveness of the detection in versatile datasets along with its computational analysis, i.e., the dataset with numerous attack vectors present, which aligns with the real-world requirements. The versatile dataset is the closest depiction of real-world scenarios with many features and a large volume of data. Apart from the fusion of datasets, it is also important to highlight that the solution for detecting DDoS attacks with versatile datasets should consider the computational capacity and prediction time. While the mentioned studies propose different deep learning-based approaches for DDoS attack detection with limited dataset attributes, further research and evaluation are necessary to improve the detection method's effectiveness, scalability, and robustness of the detection method for real-world security solutions. From the above-mentioned papers, we can summarize that there is a requirement for a solution that balances detection effectiveness, computational capacity, and prediction time in diverse and large datasets. also improving the scalability and robustness of the DDoS detection method for real-world security solutions. The main contribution of our research work is evaluating the transformer model in a versatile dataset, i.e., the combination of BoT-IoT and UNSWB15. The motivation behind evaluating the Transformer model for DDoS detection in a versatile dataset will draw the attention of the research community and the solution provider to implement it for real-time detection mechanisms.

## 3 Research Methodology

The research model implemented is depicted as a block diagram in Figure 2, consisting of several interchangeable blocks that comprise a processing pipeline. These building elements make it easier to quickly customize, train, and test our model using flow-based DDoS datasets like BoT-IoT datasets that capture IoT-specific behaviors. Each of these blocks will be covered in the sections that follow.

*3.1 Data set acquisition*

Our data ingestion component is designed to acquire datasets from various sources(e.g., NF-UNSW-NB15, NF-BoT-IoT, NF-CSE-CIC IDS2018). It is highly adaptable (As we are considering multiple datasets for our experimental setup) to any tabular data set. It seamlessly handles missing numbers or values outside the expected range. To effectively apply this component to a specific data set, a data set specification is required. This specification includes a list of categorical and numerical attributes in the data set. It also specifies the benign traffic label and the class column, which are essential for training and evaluation purposes to differentiate between malicious and legitimate flows. Once the data set specification is provided, our model can seamlessly process the data set through subsequent stages without additional code. This ensures a smooth transition from data ingestion to pre-processing and input encoding steps. By automating the data ingestion process and incorporating data set specifications, our model streamlines the handling of diverse datasets, enabling efficient data processing and analysis. There are around 43 features for pre-processing; some of the features are indicated in Table 1.

The input data set has the features in Table 1.

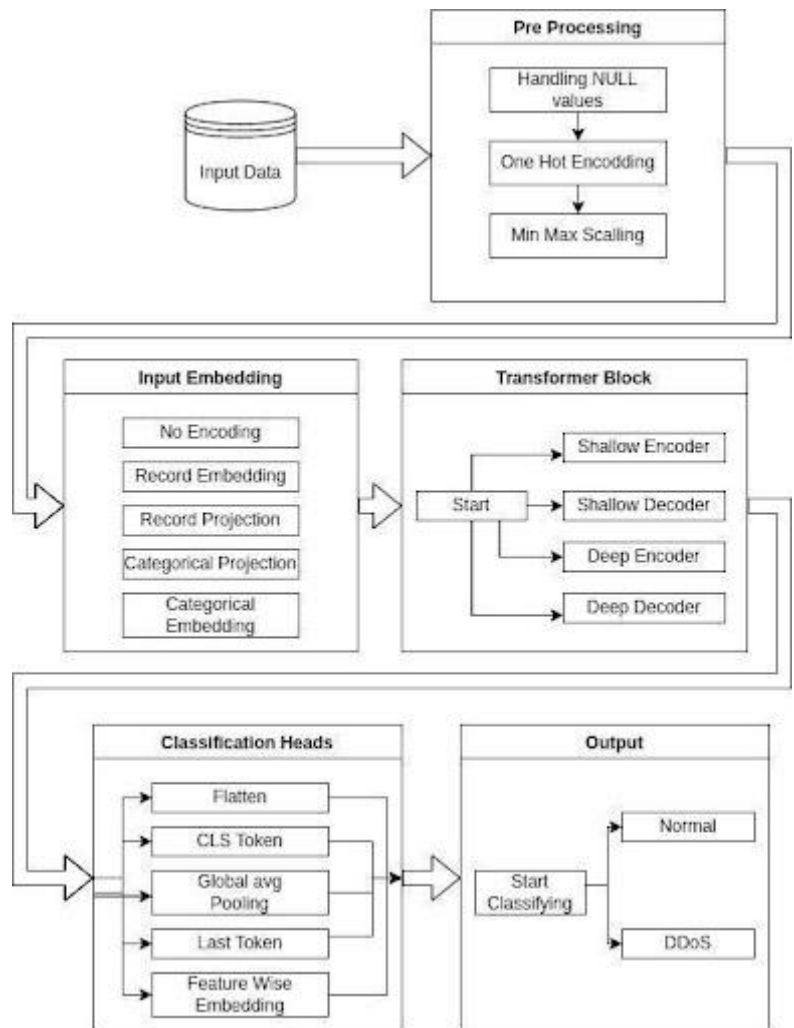

**Fig. 2**: Block Diagram of our model for detecting DDoS attacks using Transformers

*3.2 Pre-processing:*

This research used 1)NF-UNSW-NB15 and 2)NF-BoT-IoT data sets. To ensure that there is no data loss from the evaluation data set during Pre-processing, the pre-processing is first fitted to the training data and then applied to the entire data set. The data we are working with is in a flow-based format, commonly used by network managers, but requires additional processing to be suitable for machine learning models. This format includes categorical fields, such as TCP and UDP port numbers, which must be appropriately handled. For example, SSH port 22 and SMTP port 25 may have identical numerical values but represent different protocols. On the other hand, HTTPS is frequently used on ports 8080 and 443 despite having a higher numerical difference. To address these challenges, categorical data can be pre-processed using conventional methods. The pre-processing layer can either one-hot encode or integer encode the categorical variables, depending on the requirements of the input encoder. Certain machine learning models, like neural networks, benefit from representing contextual distances between values for optimal performance. In this case, pre-processing involves encoding the most frequent N categorical features using either one-hot or integer encoding and performing min-max scaling on numerical features after applying logarithm transformations. Custom pre-processing can be established using the fit and transform methods for numerical and categorical fields. The fit and transform techniques for categorical fields consider the expected categorical format (one-hot or integer encoded) as a parameter, allowing flexibility in switching between different pre-processing approaches based on the input encoding requirements. Here are the detailed changes that occur to the data after each step of pre-processing:

## Step 1: Handling Categorical Fields

• The original dataset contains categorical fields like TCP and UDP port numbers.
• For the N most frequent categorical features, we apply one-hot encoding or integer encoding based on the input encoder requirements. • One-Hot Encoding:
1. Each categorical feature with N unique categories is replaced with N binary features.
2. Each binary feature represents one category, and only one feature will have a value of 1 (indicating the presence of that category). At the same time, the rest are set to 0 (indicating the absence).
• Integer Encoding:
1. Each category is mapped to a unique integer value.
2. After this step, the categorical features are transformed into a numerical format suitable for machine learning models.

**Table 1:** Features and Meanings

| Features | Meaning |
|---|---|
| IPV4_SRC_ADDR | The source IPv4 address of the network flow. |
| L4_SRC_PORT | The source port number used in the transport layer (e.g., TCP or UDP) for the outgoing packets. |
| IPV4_DST_ADDR | The destination IPv4 address of the network flow. |
| L4_DST_PORT | The destination port number used in the transport layer (e.g., TCP or UDP) for the incoming packets. |
| PROTOCOL | The network protocol used for the flow(e.g., TCP, UDP, ICMP). |
| L7_PROTO | The Layer 7 (application layer) protocol is used for communication (e.g., HTTP, DNS, FTP). |
| IN_BYTES | The total number of incoming bytes for the flow. |
| IN_PKTS | The total number of incoming packets for the flow. |
| OUT_BYTES | The total number of outgoing bytes for the flow. |
| OUT_PKTS | The total number of outgoing packets for the flow. |
| TCP_FLAGS | The TCP flags used in the TCP header(e.g., SYN, ACK, FIN) for TCP flows. |
| ICMP_TYPE | The type of ICMP (Internet Control Message Protocol) message for ICMP flows. |
| ICMP_IPV4_TYPE | The specific type of ICMP message for IPv4 packets in ICMP flows. |
| Label | The label assigned to the network flow indicates whether it is a normal flow or an attack flow (e.g., "normal" or "attack"). |

### Step 2: Handling Numerical Features

• The original dataset contains numerical features that need pre-processing.
• We apply logarithm transformations to the numerical features. Logarithm transformations normalize the distribution and handle extreme values.
• After applying logarithm transformations, we perform min-max scaling on the numerical features. Min-max scaling scales the features to a specific range, often between 0 and 1, making them compatible with machine learning models.
• After this step, the numerical features are transformed and standardized, ready for use in training the models.

### Step 3: Training and Evaluation Split

• The preprocessed dataset is divided into training and evaluation sets.
• We allocate 80% of the data to the training set and 20% to the evaluation set.
• This ensures that the models are trained on a substantial portion of the data and evaluated on unseen data to measure their generalization performance.

The final preprocessed dataset will have the following characteristics:

• Categorical features replaced with one-hot encoding or integer encoding for the N most frequent categorical
variables. • Numerical features logarithm-transformed and min-max scaled.
• The data is split into training and evaluation sets for model development and evaluation purposes.
• This processed data is now suitable for use with machine learning algorithms, especially neural networks, to build intrusion detection models or any other relevant tasks based on the research objectives.

*3.3 Architecture of our Transformer model*

We have developed a TensorFlow Keras model of the transformer that can be independently put together and trained as required. This section is going to present the transformer model, which consists of three interchangeable parts:

*3.3.1 Input Encoding:* is crucial in converting flow data into a fixed-length feature representation, enabling effective processing by a transformer model. Before a transformer can analyze a flow, the input encoder must transform it into a meaningful feature vector[55]. Flow data, a type of tabular data, consists of predefined fields that are categorized into 2 types:
• Categorical Fields that contain discrete levels, such as port numbers. However, numerical proximity doesn't always imply contextual similarity. For example, ports 22 and 25 are numerically close but serve different protocols, whereas ports 443 and 8080 are widely used for similar services despite being numerically distant[55].
• Numerical Fields includes metrics such as the number of packets sent, where the distance between values carries contextual significance[55].

Different encoding strategies ensure that the input encoder captures categorical and numerical data. 1 commonly used method, as in *TabTransformer*, encodes categorical fields individually by converting them into continuous vectors that preserve contextual meaning. These vectors are combined with numerical fields to generate feature vectors suitable for machine learning models. However, this approach primarily focuses on encoding categorical information, treating numerical fields separately. Another approach involves embedding categorical and numerical fields simultaneously. This is achieved by applying one-hot encoding to categorical fields, concatenating them with numerical values, and passing the combined data through an encoding layer. While this method enables comprehensive analysis, it can lead to high-dimensional feature spaces, making certain fields more challenging to encode effectively[55]. By selecting an appropriate encoding strategy, the input encoder ensures that the feature vectors retain essential information, allowing the transformer to process the flow data accurately[55]. Here's a brief description of each component of a transformer model:

Encoder:

• The encoder takes an input vector sequence consisting of multiple layers (L layers).
• Each layer in the encoder contains a self-attention mechanism and a feed-forward neural network.
• The self-attention mechanism computes weighted sums of input vectors to generate contextualized representations. • The feed-forward neural network introduces non-linearities and further processes the information from self-attention. • The encoder processes the input sequence independently and generates encoded representations for each position in the sequence. Output of each sublayer = LayerNorm(x + Sublayer(x))

Self-Attention Mechanism:

• The self-attention mechanism computes attention scores between each vector in the input sequence (query Q) and all other vectors (keys K and values V), which can be computed using Equation 1.
• Query, key, and value projections are performed using learnable weight matrices.
• Attention scores are obtained by taking the dot product of the query and key projections.
• Attention weights are obtained by applying the softmax function to the attention scores.
• The weighted sum of the value vectors, using the attention weights, produces the output of the self-attention mechanism.
!

$$\text{Attention}(Q, K, V) = \text{softmax}\left(\frac{QK^T}{\sqrt{d_k}}\right)V \quad (1)$$

Feed-Forward Neural Network (FFN):

• The feed-forward neural network consists of two linear transformations with a non-linear activation function in between. It is calculated using Equation 2.
• It introduces non-linearities to the output of the self-attention mechanism, enhancing the model's ability to capture complex patterns in the data.

$$FFN(x) = \max(0, xW_1 + b_1) W_2 + b_2 \quad (2)$$

Decoder:

• The decoder, like the encoder, consists of multiple layers (L layers).
• Each layer in the decoder contains a self-attention mechanism, an encoder-decoder attention mechanism, and a feed-forward neural network. • The decoder generates the output sequence based on the encoded representations produced by the encoder.

Encoder-Decoder Attention:

• The encoder-decoder attention mechanism allows the decoder to attend to the encoder's output, incorporating information from the input sequence.
• It computes attention scores between the decoder's query and the encoder's keys, allowing the decoder to focus on relevant parts of the input.

In summary, a transformer model uses the encoder to process the input sequence independently, producing encoded representations. The decoder then generates the output sequence, leveraging the encoded representations and attending to the input sequence when necessary. The self-attention mechanism and feed-forward neural networks are the core components that capture dependencies and non-linear relationships within the data, making transformers effective in various natural language processing tasks.

*3.3.2 Classification Head Options:* Transformers are sequence-to-sequence models; therefore, we must transform the sequential output into a classification result to accomplish our goal. We have identified alternate methods that can be more effective than directly feeding all the transformer outputs into the MultiLayer Perceptron (MLP) for classification. One often employed technique is "flattening, " which converts the output sequence into a feature vector before being fed into the MLP. However, because the number of parameters is increasing exponentially, this strategy gets less effective as the sequence length increases. We have investigated further approaches, such as global average pooling, to overcome this problem. To do this, a feature vector is created by averaging the features in the output sequence. Global average pooling works effectively for tasks like sentiment classification, where it is advantageous to average the characteristics throughout the full phrase or paragraph. Averaging over the context of earlier flows may not be the best option when just the last flow in the sequence is being classified. Another strategy we consider is using a time-distributed neural network layer for each feature in the contextual representation. This enables us to represent the last flow's additional weight by applying different weights to the other flows in the sequence. This strategy, also known as feature-wise embedding or feature-wise projection, avoids including data from earlier flows that the transformer has already considered. Alternatively, we can enter the feature vector for the classification head from the transformer's final output. This avoids the need for averaging or additional layers and corresponds to the feature vector of the previous flow. This strategy is known as the "Last token."We also investigate the effectiveness of adding a "classification" token at the end of the series of flows. This strategy is comparable to models like BERT, where a unique token directs the model to carry out a certain task. We assess the effectiveness of our token-based strategy by classifying data using the corresponding transformer output. In conclusion, we explore various methods for converting a transformer model's sequential output into a classification outcome. These techniques increase efficiency and optimize parameter counts for various activities, including flow classification.

*3.4 Details of the dataset and learning model*

The dataset that we utilized to train our model and the outcomes we obtained are covered in this chapter. We used a grid search to gather the results. Input encodings, transformer block type, depth, transformer feed-forward size, attention heads per transformer, classification heads per transformer, and learning rate are the dimensions investigated in this article.

*3.4.1 Model Training and Grid Search:* Each experiment has been tried at least three times when conducting a grid search over a target space and opted for the best repeat, which can account for insufficient model initialization throughout the experimentation, to determine the outcome. Accordingly, the early halting and an epoch limit for the model training have been set. The maximum number of epochs was 20, and the early halting was set to 5 epochs of patience. Since most models had converged to within 1% of their final performance by the 20th epoch during our initial experimentation, we decided to use 20 epochs. With the learning rate set in each experiment, we trained using the Adam optimizer. Our model was run on Google Colab and had specifications for Intel Xeon CPU @2.20 GHz, 13 GB RAM, Tesla K80 accelerator, and 12 GB GDDR5 VRAM.

*3.4.2 Datasets:* In this research, two prominent datasets for DDoS attacks have been used, i.e.

1. The University of New South Wales produced the network security dataset known as UNSW-NB15-V2
(https://research.unsw.edu.au/projects/unsw-nb15-dataset).
It includes legitimate and nefarious information on network activity, such as DoS and DDoS attacks. It is used to train and test network security algorithms and intrusion detection systems and has over 2 million instances. The dataset is useful for enhancing network security measures since it allows researchers to analyze and categorize various kinds of network behavior.

**Table 2** Class Distribution in UNSW-NB15 dataset

| Class | No of records | % of total data |
|---|---|---|
| Benign | 2295222 | 96.023 |
| Exploits | 31551 | 1.32 |
| Fuzzers | 22310 | 0.933 |
| Generic | 16560 | 0.69 |
| Reconnaissance | 12779 | 0.534 |
| DoS | 5794 | 0.242 |
| Analysis | 2299 | 0.096 |
| Backdoor | 2169 | 0.090 |
| Shellcode | 1427 | 0.059 |
| Worms | 164 | 0.006 |

2. UNSW Canberra's Cyber Range Lab created the BoT-IoT dataset and offers a complete network environment, including botnet and regular traffic (https://research.unsw.edu.au/projects/bot-iot-dataset).
It is divided into attack kinds and subcategories and contains a variety of file formats, including pcap, argus, and csv. The dataset includes attacks such as data exfiltration, keylogging, OS and service scans, DDoS, and DoS. A 5% subset that totals 1.07 GB and 3 million records has been extracted. The BoT-IoT dataset is important for researching and creating secure IoT settings.

**Table 3:** Class distribution in BoT-IoT dataset

| Class | No of records | % of total data |
|---|---|---|
| Non anomalous(normal) | 9543 | 0.013 |
| Information gathering | 1821639 | 2.480 |
| DDoS | 38532480 | 52.500 |
| DoS | 33005194 | 45.00 |
| Information theft | 1587 | 0.002 |

*3.4.3 Evaluation Metrics:* Standard criteria were used to evaluate the effectiveness of various transformer models, including the false alarm rate, detection rate, and F1 score. True Positives, True Negatives, False Positives, and False Negatives—abbreviated as TP, TN, FP, and FN are combined to calculate the metrics.

*3.4.4 Hyperparameters:* Different input encoding techniques, classification heads, transformer configurations, and corresponding hyperparameters were examined. We took into account the following hyperparameters for our model:

• Transformer Block Types: The encoder and decoder blocks have been considered in the proposed model setup. • Number of Layers: To investigate the effects of varied layer counts on model performance, the experiments were conducted with 2, 4, 6, and 8 layers.
• Feed-forward Dimensions (FF): To find the best size for the model range of feed-forward layer sizes in the transformer blocks, 128, 256, and 512 have been chosen.
• Number of Attention Heads: From 2 to 12 to analyze the impact of various attention head counts in the transformer blocks. • Learning Rate: To determine the best value for the model's training, the range of learning rates was 0.01, 0.001, 0.0005, 0.0001, and 0.00001.
• Optimizer: The model was optimized using the Adam optimizer, known for its efficiency in handling sparse gradients on noisy problems. The binary cross-entropy loss function was employed, which was appropriate for binary classification tasks, ensuring that the model's predictions were as close to the actual labels as possible.
• Batch size: During training, a batch size of 128 was used, striking a balance between training speed and memory efficiency. • Epochs: The model was trained for a maximum of 20 epochs, with each epoch comprising 64 steps per epoch. • Early stopping: To prevent overfitting and enhance the model's ability to generalize to unseen data, early stopping was implemented with a patience of 5 epochs.

Various hyperparameters and configurations were evaluated to better understand how they affect the model's performance and find the right set for our particular purpose.

# 4 Results and Discussion

In this section, we present the results of our approach in two parts. The efficacy of the transformer model is demonstrated by comparing two models, Transformer and LSTM, furthering the computational analysis of the system.

*4.1 Efficacy of the system*

Various parameters, e.g., Input Encoding, Classification Heads, different Transformer models, etc., affected the model's overall performance. Figure 3 shows the confusion matrix of the Transformer model.

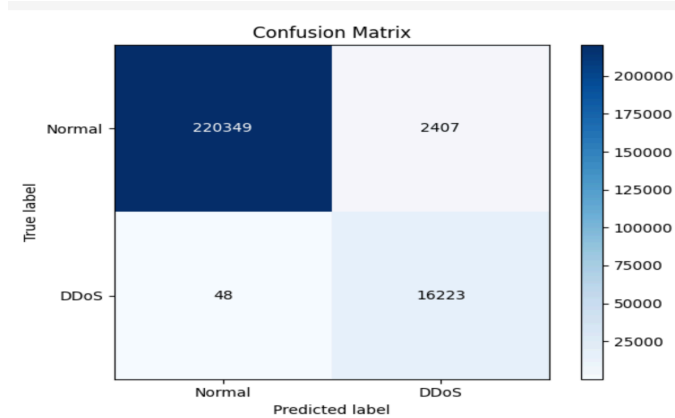

**Fig. 3**: Confusion Matrix

Table 4 compares the performance of the four transformer models we considered for our experimental setup. Here, we have shown the two best-performing combinations for each transformer model.

$$\text{F1 Score} = \frac{2TP}{2TP + FP + FN} \quad (3)$$

It is evident from Table 4 that the type of transformer model didn't have much effect on the F1 score and accuracy of the model. The F1 score is calculated using Equation 3. Though, to be quite precise, we can say the shallow models outperformed the deeper model when we check their F1 score and accuracy. Though their performances were insignificant, the major difference we noticed between the shallow and deeper model was in terms of their total trainable parameters, which can be seen in Table 4. The deep encoder and decoder models, like GPT and BERT, had a much larger number of trainable parameters than the shallow models. Its major effect was on the time required for training the models and their inference time. For a resource-constrained environment, such as an IoT environment, deep encoder and decoder models will pose a huge problem. As our basic objective of this research work was to identify a lightweight machine learning model that can be implemented for

**Table 4** Comparison of the transformer models

| Transformer | Layers | Heads | Dim | Classification head | Input encodings | Parameters | F1 score | Accuracy | Alarm Rate |
|---|---|---|---|---|---|---|---|---|---|
| Shallow Encoder | 2 | 2 | 128 | Last Token | Record Emb.Dense | 383,237 | 90.54 | 99.79 | 1.05 |
| Shallow Encoder | 2 | 2 | 256 | Feature Wise Projection | Record Projection | 349,332 | 90.43 | 98.38 | 0.95 |
| Shallow Decoder | 2 | 2 | 128 | Last Token | No input encoding | 1,346,545 | 88.74 | 99.65 | 1.32 |
| Shallow Decoder | 2 | 2 | 128 | Flatten | No input encoding | 1,568,666 | 87.81 | 99.53 | 1.12 |
| GPT Model(Deep Decoder) | 12 | 12 | 768 | Feature wiseprojection | Categorial Emb. Lookup | 20,514,132 8 | 87.89 | 98.62 | 1.06 |
| GPT Model(Deep | 12 | 12 | 768 | Last Token | Record Emb.Dense | 30,506,179 8 | 89.93 | 98.8 | 1.05 |

| | | | | | | | | |
|---|---|---|---|---|---|---|---|---|
| Decoder) | | | | | | | | |
| BERT Model(Deep Encoder) | 12 | 12 | 768 | Last Token | No input encoding | 124,875 | 87.79 | 98.75 | 1.23 |
| BERT Model(Deep Encoder) | 12 | 12 | 768 | Last Token | Record Projection | 29,862,401 | 76.32 | 98.90 | 2.78 |

In real-time scenarios, it's apparent that BERT and GPT won't be a good choice for such use cases. This is well-suited for DDoS attack detection tasks where we have a real-time stream of network traffic packets and must monitor them continuously to detect the attack. It can be helpful to find out if the recent flow is part of a suspicious sequence. Table 4 shows that the input encoding did not impact the overall model's F1 score. The performance of each input encoding technique was essentially the same. However, it significantly reduced the number of model parameters. For instance, record-level embedding used the last token as the classification head and produced the same outcome as category embedding with only half the parameter counts.

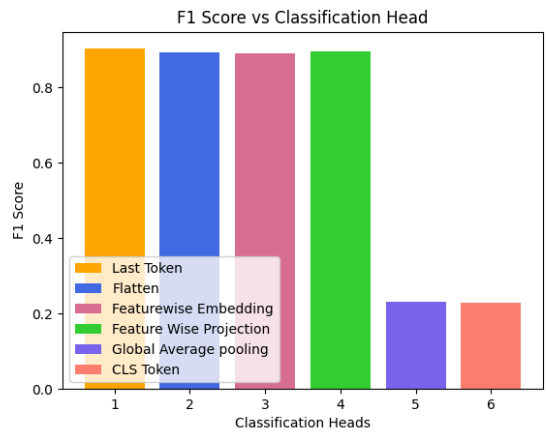

**Fig. 4**: F1 scores obtained for different types of classification Heads

It can be observed from Figure 4 that classification heads have the biggest impact on the model's performance. While global average pooling and CLS token did, Last Token, Flatten, Feature-wise Embedding, and Feature-wise Projection all performed roughly the same, and with satisfactory results. The authors of most NLP research papers have employed global average pooling, although it should be avoided when doing DDoS detection tasks. This can be considered as one of the major contributions. Global Average Pooling has been used extensively in related areas, particularly natural language processing, to accomplish categorization. However, our research work's outcomes showed this was one of the least effective techniques. This makes sense, given that only the last flow in the sequence is being classed and that the transformer previously captured the relationships between this flow and earlier flows in the previous token. By averaging, we incorporate the data from earlier flows that might not be pertinent to the current classification task. The Flatten method was one of the best classification heads, which takes all of the transformer's outputs and sends them straight to the classification Multi-Layer Perceptron(MLP). Figure 4 reveals that, in contrast to the other approaches, these results in a disproportionately high parameter count that grows exponentially with sequence length. Although this method can be employed for short flow sequences, the model size may grow out of control for longer sequences. More training data are needed, and the risk of over-fitting grows with more model parameters. Finally, feature-wise embedding also had decent performance. With this method, the classification multi-layer perceptron can get a weighted mixture of all the features from the flows in the sequence instead of simply the data from the previous flow. As the transformer has already recorded dependencies between the latest flow and earlier flows, this is probably not preferable to taking only the last flow's embedding. Considering that this receives a slightly lower F1 score than picking the last token, it follows that final token classification heads seem to be the most effective for DDoS attack detection tasks.

**Table 5** Training Vs Inference Time

| Transformer model | Classification Head | Input Encoding | Train time(flows/sec) | Inference Time |
|---|---|---|---|---|
| Shallow Encoder | Last Token | Record Emb. Dense | 1630 | 7877 |

| Shallow Encoder | Feature Wise Projection | Record Projection | 1527 | 6321 |
| Shallow Decoder | Last Token | No input encoding | 1095 | 6481 |
| Shallow Decoder | Flatten | No input encoding | 1088 | 6321 |
| GPT Model (DeepDecoder) | Feature Wise Projection | Categorical Emb. Lookup | 105 | 287 |
| GPT Model (DeepDecoder) | Last Token | Record Emb. Dense | 288 | 2959 |
| BERT Model (DeepEncoder) | Last Token | No input encoding | 48 | 154 |
| BERT Model (DeepEncoder) | Last Token | Record Projection | 83 | 752 |

In Table 5, train time is the duration for a machine learning model to learn patterns in a dataset, while inference time is the phase where the trained model predicts or performs tasks on new data, which is crucial for real-time systems with low latency. They are measured in terms of flows/second. This means that the higher the throughput (flows/second), the better the model will be to train, and it will take less time for prediction. A model with higher throughput will be preferred for the real-time implementation of our model. The shallow transformers have much higher throughput than the deeper models. To measure the training and inference time, we followed the following steps:

Training Time Measurement:

• The GPU was cleared, and TensorFlow was warm before starting the measurements.
• The training time for each batch was recorded, including the backpropagation step.
• The recorded training time for each batch was divided by the batch size.
• The average training time per batch was calculated by averaging all the batches used during training.
• TensorFlow train on batch function was used for this purpose.
• Care was taken to ensure no other processes were running during the model training.
• Outlier testing was performed to check for significant drift in batch training times during training.

Inference Time Measurement:

• The inference time for a single batch was recorded from the start to the end of the batch.
• This process was repeated four times on the same batch of data to ensure consistent results and avoid caching effects. • The median time from the four measurements was taken to obtain a stable range for inference time.
• This procedure was repeated for 50 batches of data, selected randomly.
• The mean time over all the batches was calculated to determine the average inference time.

Finally, the result of the proposed research is being compared with the state-of-the-art method, where LSTM was used as the classifier.

**Table 6** Comparison of our model with state-of-the-art models

| Technique | Algorithm | Data-set | No of Features | Amount of dataset used | Accuracy | No of Classes Classified |
|---|---|---|---|---|---|---|
| Protocol-Based Deep Intrusion Detection for DoS and DDoS Attacks. | LSTM | BoT-IoT UNSW NB15 | 26 | 96% of the BoT-IoT dataset and 87% of the UNSW NB15 dataset | 96.3 | 3 |
| A Transformer Based Approach for DDoS Attack Detection in IoT Networks (Proposed). | Shallow Encoder, Shallow Decoder, Deep Encoder, Deep Decoder | BoT-IoT, UNSW NB15 | NA | Used the entire dataset, out of which 80% was used for training and 20% for testing | 99.79, 99.65,98.62, 98.75 | 2 |

*4.2 Computational Analysis*

CPU utilization (per core) over time:
The consistency indicates that the transformer model is likely processing at a steady rate without significant spikes or drops in CPU load, which implies a stable processing demand. The Transformer model appears to have a more consistent and possibly lower computational demand than the LSTM model.

Disk utilization in gigabytes (GB) over time:
The disk utilization appears to be constant at around 27 GB, indicating that the Transformer model uses no additional disk space during this time interval. The Transformer model does not show any disk space change over the 30 seconds depicted, which means the model is loaded in memory and not engaging in significant read/write activities during this period. The LSTM model has a small but noticeable increase in disk space usage at the start, reflecting initialization activities or initial data processing steps. After this initial activity, the disk usage does not change, suggesting that the model is either running in memory or that data is not being written to or read from the disk.

**Table 7** Comparison of Computational Analysis

| Parameters | LSTM | Transformer | Difference | Inference |
|---|---|---|---|---|
| CPU utilization (percore) | 90% | 30% | 60% | Transformer model has CPU utilization 60% less than LSTM model. |
| Network traffic(bytes) | 0-50MB | 0.4MB-1MB | 49MB | Network traffic drops by 98% for Transformer model. |
| Disk utilization (GB) | 27.4GB | 27GB | 0.4GB | Even though the difference is small, the duration of disk utilization for the LSTM model is in minutes, whereas for the Trans former model, it is in seconds. |

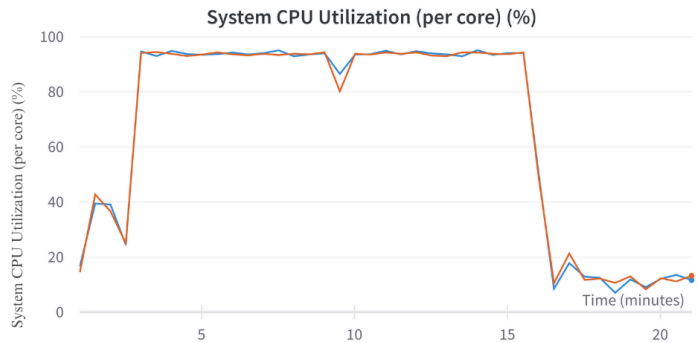

**Fig. 5**: System CPU utilization for LSTM

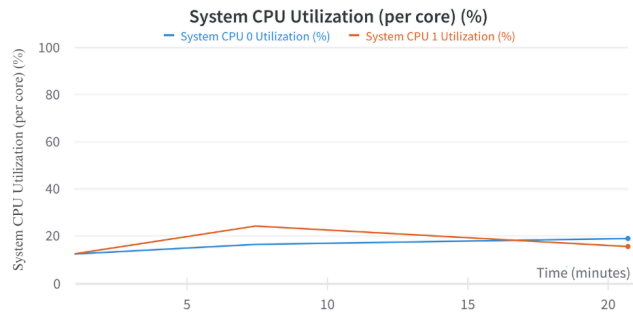

**Fig. 6**: System CPU utilization for Transformer

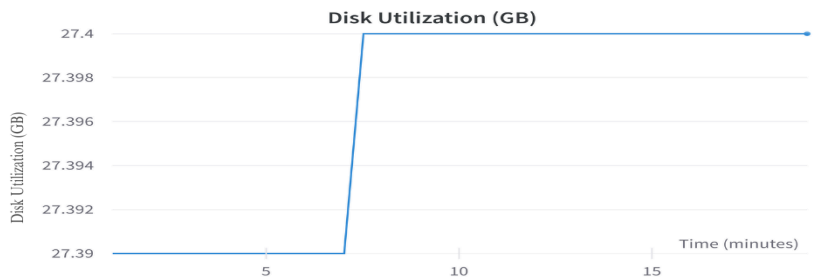

**Fig. 7**: Disk utilization for LSTM

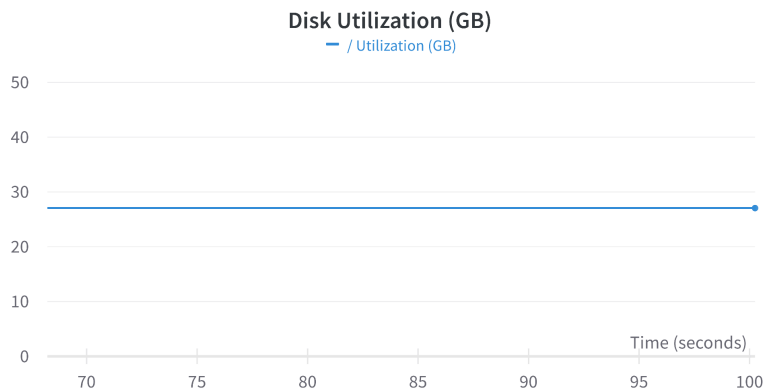

**Fig. 8**: Disk utilization for Transformer

Network traffic in bytes over time:
The graph indicates that the amount of network traffic associated with the LSTM model also increases as time progresses. We can infer that both models engage in network activity, with the LSTM model showing a larger total traffic volume over a longer period. In contrast, the Transformer model shows a detailed, shorter, balanced two-way network traffic period. As the LSTM model has 2 LSTM layers, one for binary classification and a second for multi-class classification, along with a dense layer for output and an embedded layer for converting the dataset to a vector, the network traffic is more compared to the Transformer model.

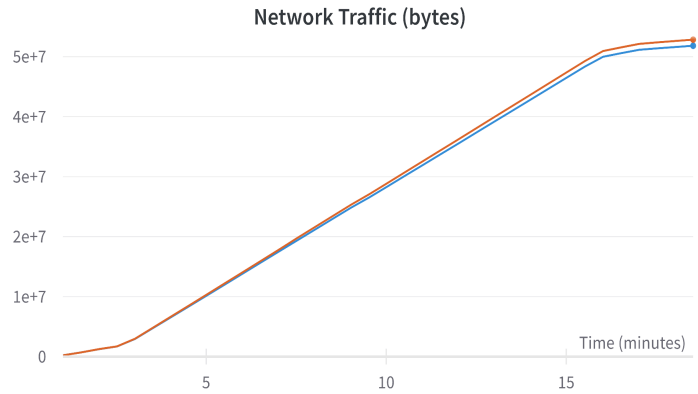

**Fig. 9**: Network traffic for LSTM

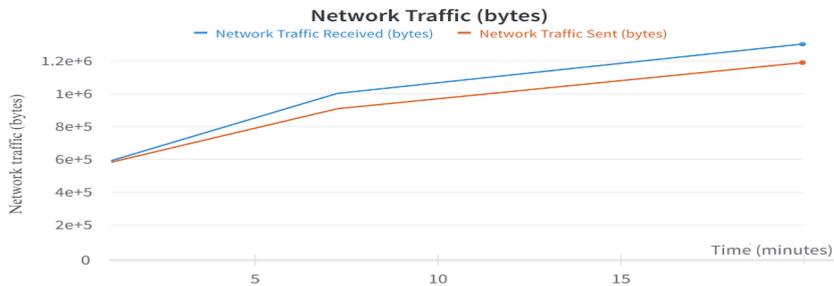

**Fig. 10**: Network traffic for Transformer

GPU Power Usage (W):
GPU power usage, which tracks the power consumption of a GPU (Graphics Processing Unit) over time, measured in seconds. The horizontal x-axis represents the time from 70 to 100 seconds, while the vertical y-axis represents the power usage in watts, ranging from approximately 11.75W to 12.05W. The curve shows a slight bend, indicating a gradual acceleration in power usage rather than a linear increase. The increase in power usage suggests that the GPU is being subjected to progressively more demanding tasks, such as computational processes, that are becoming more intensive over time.

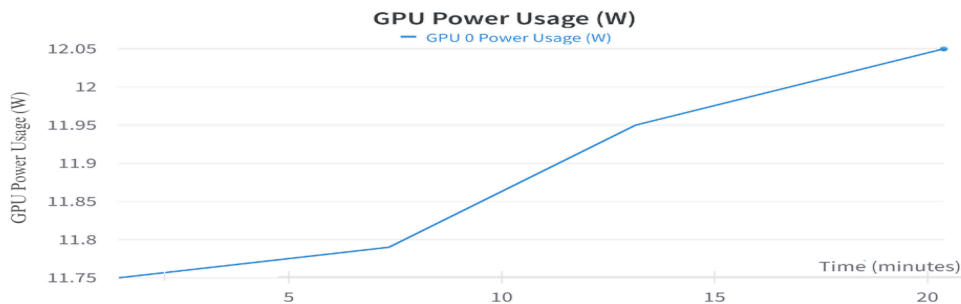

**Fig. 11**: GPU power utilization

## 5 Conclusion

The proposed research has demonstrated the utility and power of transformer models by a detailed and methodical analysis of various configurations over a combination of two widely used datasets. With the ability to process large amounts of network traffic data in real-time and identify anomalous patterns, Transformer models can provide accurate and timely detection of DDoS attacks. Through self-attention mechanisms, Transformer models can capture long-range dependencies in the data and learn contextual relationships, making them particularly effective in detecting sophisticated and evolving DDoS attacks when working with such enormous data. Furthermore, adapting to changing network traffic patterns and learning from new data ensures these models remain effective. Our research concluded that selecting the classification head was the most important aspect of the model's performance, and "Last Token" was the best option. The model size can be decreased by more than half without affecting the classification performance by applying record-level embedding or projection. The quickest approaches in terms of training time and inference were record-level approaches. For DDoS detection tasks, shallow transformer models are adequate, achieving the highest accuracy of 99.79% with 60% less CPU utilization as compared to the state-of-the-art LSTM model, which had an accuracy of 96.3%, as shown in Table 6.

## 6 Future Work

In light of the challenges identified with our proposed approach, future work can be focused on several key areas to enhance the robustness and effectiveness of our DDoS detection system in IoT networks. First, a mechanism for incremental learning can be developed, enabling the model to continuously learn from new data without needing to be retrained from scratch. This will help the model adapt to evolving network conditions and data patterns, ensuring it remains effective. Additionally, incorporating adversarial training techniques will make the model more resilient against evasion attacks, where attackers manipulate traffic patterns to avoid detection. To address the practical challenges of implementing and maintaining detection systems, automated tools can be developed for model monitoring, updating, and retraining, reducing the need for extensive manual intervention. Enhancement of the system's real-time data handling capabilities to preprocess and analyze network traffic efficiently, meeting the stringent latency requirements of IoT networks. Extensive real-world testing of the model in various IoT environments can be conducted to validate its effectiveness and identify areas for improvement. The model can be integrated into existing IoT frameworks at the Modular level as part of security pipelines, such as IDS and SIEM systems, or at the edge or cloud level, leveraging the computational power of cloud resources for large-scale data analysis. Establish benchmarks for different types of IoT networks and attack scenarios, systematically evaluating the model's performance and guiding future enhancements. These efforts will collectively ensure that our DDoS detection system remains robust, adaptive, and effective in real-world IoT environments.

## 7 Data availability statement

The following datasets were used for the experimental work; the links have been provided for reference:

1. UNSW-NB15: https://research.unsw.edu.au/projects/unsw-nb15-dataset
2. BoT-IoT: https://research.unsw.edu.au/projects/bot-iot-dataset